\documentclass[pra,10pt, twocolumn, floatfix, superscriptaddress]{revtex4-1}
\usepackage{amsmath} % need for subequations
\usepackage{graphicx} % need for figures
\usepackage{verbatim} % useful for program listings
\usepackage{color} % use if color is used in text
\usepackage{subfigure} % use for side-by-side figures
\usepackage{hyperref} % use for hypertext links, including those to external documents and
\usepackage{sidecap}
\usepackage{tikz}
\usepackage{amsmath,bm}
\usepackage{amsfonts}
\usepackage{mathtools}
\usepackage{epstopdf}
\usepackage{float}
\usepackage{physics}

\begin{document}

%\title{Decay of a quantum knot into a polar-core spin vortex in a Bose--Einstein condensate}
\title{Decay of a Quantum Knot}
\author{T. Ollikainen}
\email{tuomas.ollikainen@aalto.fi}
\affiliation{QCD Labs, QTF Centre of Excellence, Department of Applied Physics, Aalto University, P.O. Box 13500, FI-00076 Aalto, Finland}
\affiliation{Department of Physics and Astronomy, Amherst College, Amherst, MA 01002–5000, USA}
\author{A. Blinova}
\affiliation{Department of Physics, University of Massachusetts, Amherst, MA 01003, USA}
\affiliation{Department of Physics and Astronomy, Amherst College, Amherst, MA 01002–5000, USA}
\author{M. M\"ott\"onen}
\affiliation{QCD Labs, QTF Centre of Excellence, Department of Applied Physics, Aalto University, P.O. Box 13500, FI-00076 Aalto, Finland}
\affiliation{VTT Technical Research Centre of Finland Ltd, P.O. Box 1000, FI-02044 VTT, Finland}
\author{D. S. Hall}
\affiliation{Department of Physics and Astronomy, Amherst College, Amherst, MA 01002–5000, USA}


\begin{abstract}
We experimentally study the dynamics of quantum knots in a uniform magnetic field in \mbox{spin-1} Bose--Einstein condensates. The knot is created in the polar magnetic phase, which rapidly undergoes a transition towards the ferromagnetic phase in the presence of the knot. The magnetic order becomes scrambled as the system evolves, and the knot disappears. Strikingly, over long evolution times, the knot decays into a polar-core spin vortex, which is a member of a class of singular $\mathrm{SO(3)}$ vortices. The polar-core spin vortex is stable with an observed lifetime comparable to that of the condensate itself. The structure is similar to that predicted to appear in the evolution of an isolated monopole defect, suggesting a possible universality in the observed topological transition. 
\end{abstract}

\maketitle

Topological defects and textures provide intriguing conceptual links between many otherwise distant branches of science~\cite{Mermin:1979,Nakahara:2003}. They appear in various contexts ranging from condensed matter to high-energy physics and cosmology, and can be highly stable against weak perturbations. However, there can be mechanisms leading to the decay of the defects despite their topological stability. The decay can be induced by, for example, changes to the underlying symmetries or the finite size of the system~\cite{Preskill:1993}. 

Spinor Bose--Einstein condensates (BECs) are one of the most fascinating systems available for the study of topological defects due to the diverse range of broken symmetries associated with the different magnetic phases of the system. In the scalar case, the spin degrees of freedom are inaccessible and the topology of the BEC is simply described by the broken $\mathrm{U(1)}$ symmetry, yielding one-dimensional solitons and vortex lines as the only possible topological defects of the system. Upon including the spin degrees of freedom, the internal symmetries of the gas become plentiful, allowing for a diverse set of excitations. For example, in spinor BECs there can be several types of vortices~\cite{Matthews:1999,Leanhardt:2002,Leanhardt:2003,Sadler:2006,Donadello:2014,Seo:2015}, skyrmions~\cite{Ruostekoski:2001,Khawaja:2001,Savage:2003_2,Choi:2012,Lee:2018}, monopoles~\cite{Pietila:2009,Ray:2014,Ray:2015,Sugawa:2016,Ollikainen:2017_2}, and quantum knots~\cite{Kawaguchi:2008,Hall:2016}.

Topologically stable knots are classified by a linking number (or Hopf charge) $\mathcal{Q}$, which counts the number of times each preimage loop of the order parameter is linked with every other such loop~\cite{Hopf:1931}. In Ref.~\cite{Hall:2016}, the experimental creation of knots with $\mathcal{Q}=1$ was reported in the polar magnetic phase of spin-1 BECs. Alternative methods to create knots were theoretically proposed in Refs.~\cite{Ollikainen:2017,Liu:2019}. During its evolution, the knot is predicted to facilitate the decay of the underlying polar magnetic phase into the ferromagnetic phase~\cite{Kawaguchi:2008}. Prior to the present study, however, neither this nor any other prediction involving the temporal evolution of the knot has been experimentally tested beyond the preliminary investigations of Ref.~\cite{Hall:2016}.

In this Letter, we report experimental observations of the evolution of the quantum knot in spin-1 $^{87}\mathrm{Rb}$ BECs in a uniform external magnetic field. We show that the knot structure begins to decay rapidly on a time scale of several milliseconds. During the decay process, the underlying polar magnetic phase is partly replaced by the ferromagnetic phase. For long evolution times, on the order of seconds, the knot is completely destroyed and we observe a spatial rearrangement of magnetic phases, such that the polar phase occupies the central region of the condensate, surrounded by a mixed-phase region that approaches the ferromagnetic phase at the condensate boundary. Quite surprisingly, this emergent texture is that of a singular polar-core spin vortex~\cite{Saito:2006,Sadler:2006,Weiss:2019}. It begins to emerge spontaneously in the first $500~\mathrm{ms}$ of evolution and the long lifetime of the polar-core spin vortex suggests that it is a stable excitation under these experimental conditions. 

\begin{figure}[t]
\includegraphics[width=0.5\textwidth]{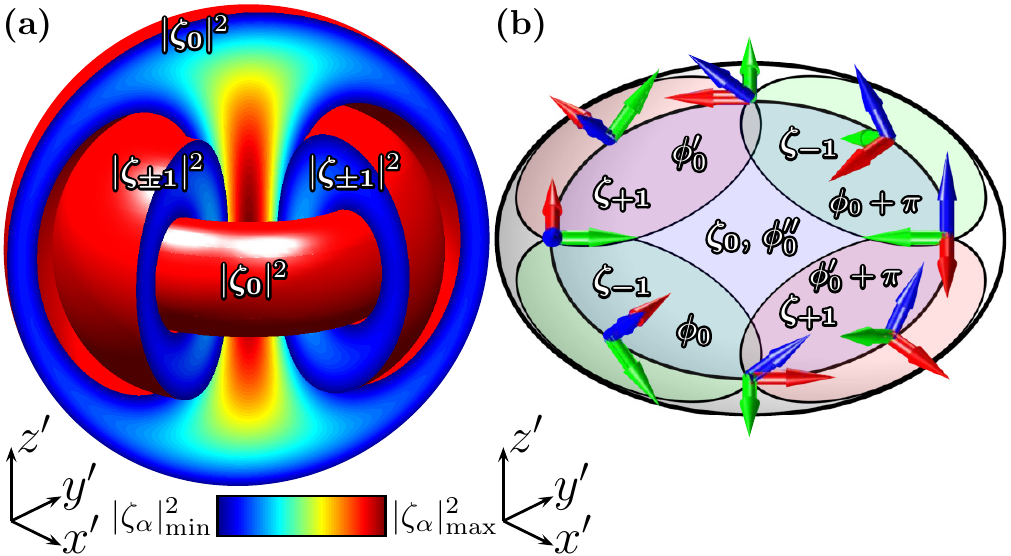}
\caption{(a) Schematic representation of isosurfaces and densities in the different spinor components of the quantum knot in the scaled coordinate system $(x',y',z')=(x,y,2z)$. The densities are revealed by partially cutting the regions in the spinor components. The red-colored isosurfaces and the color gradient minimum (dark blue) correspond to the value $|\zeta_\alpha|^2_\mathrm{min}=0.29$ while the maximum gradient color (dark red) to $|\zeta_\alpha|^2_\mathrm{max}=1$. (b) Schematic representation of the cylindrical spinor structure of the polar-core spin vortex in the $z=0$ plane resulting from a long-time evolution of the quantum knot. The red, blue, and green arrows in the triads represent ${\bf m}$, ${\bf n}$, and ${\bf s}$ vectors, respectively. The large grey circle denotes the region with nonvanishing total particle density, whereas red, blue, and green circles enclose the regions in which the predominant spinor component is $\zeta_{+1}$, $\zeta_{0}$, and $\zeta_{-1}$, respectively. The $\phi_0$,
$\phi_0'$, and $\phi_0''$ are the reference regional phases of $\zeta_{-1}$, $\zeta_{+1}$, and $\zeta_{0}$, respectively, showing the $\pi$ phase difference of the adjacent regions in the $\zeta_{\pm 1}$ components. For this schematic, we choose the phases $\phi_0=\pi/2$, $\phi_0'=0$, and $\phi_0''=\pi/2$.}\label{fig:schematics}
\end{figure}

\emph{Methods.}---An accurate and convenient description of the zero-temperature dilute spin-1 BEC is given by the mean-field theory. Within this formalism, the condensed gas is described by an order parameter which in the $z$-quantized spin basis $\{\ket{+1},\ket{0},\ket{-1}\}$ reads $\Psi({\bf r},t)=(\psi_{+1}({\bf r},t),\psi_{0}({\bf r},t),\psi_{-1}({\bf r},t))^\mathrm{T}_z$, where $\psi_\alpha({\bf r},t)=\sqrt{n({\bf r},t)}\exp[i\phi({\bf r},t)]\zeta_\alpha({\bf r},t)$, $n$ is the particle density, $\phi$ is the condensate phase, and $\zeta_\alpha$ is the spinor component with the spin quantum number $\alpha\in\{-1,0,1\}$. The spinor $\zeta=(\zeta_{+1},\zeta_{0},\zeta_{-1})^\mathrm{T}_z$ satisfies $\zeta^\dagger\zeta=1$. The dynamics of the mean-field order parameter are determined by solving the Gross--Pitaevskii equation (see Supplemental material for numerical methods). %(see Supplemental material at [URL will be inserted by publisher] for numerical methods.)

A transformation into a Cartesian basis, $\zeta_x=(-\zeta_{+1}+\zeta_{-1})/\sqrt{2}$, $\zeta_y=-i(\zeta_{+1}+\zeta_{-1})/\sqrt{2}$, and $\zeta_z=\zeta_0$~\cite{Kawaguchi:2012}, gives rise to two real-valued vectors, ${\bf m}$ and ${\bf n}$. The components of these vectors are defined through the relation $\zeta_a=(m_a+i n_a)/\sqrt{2}$, with $a\in\{x,y,z\}$. 

In the pure ferromagnetic phase, ${\bf m}$ and ${\bf n}$ are orthogonal and give rise to the spin vector ${\bf s}={\bf m}\times{\bf n}$. The orthonormal triad $(\hat{\bf m},\hat{\bf n},\hat{\bf s})$ can thus be used to describe the pure ferromagnetic order parameter, with the configuration space homotopic to $\mathrm{SO(3)}$. In the pure polar magnetic phase the spin vanishes and the Cartesian representation of the order parameter can be expressed as $\Psi({\bf r},t)=\sqrt{n({\bf r},t)}\exp[{i\phi({\bf r},t)}]\hat{\bf d}({\bf r},t)$, where $\hat{\bf d}=(d_x,d_y,d_z)^\mathrm{T}$ is a real-valued vector describing the nematic orientation in the condensate. In this case, by definition, ${\bf n}$ vanishes and ${\bf m}$ is parallel to $\hat{\bf d}$. %With a suitable choice of gauge $\exp(i\phi)$ one can make ${\bf m}$ or ${\bf n}$ parallel to $\hat{\bf d}$. 

The condensate can also reside in a mixed state where polar and ferromagnetic phases coexist, and the nematic director $\hat{\bf d}$ and the spin ${\bf s}$ are simultaneously well defined. The director $\hat{\bf d}$ can be extracted from the magnetic quadrupole moment matrix $Q$, defined in the Cartesian basis through $Q_{ab}=(\zeta_a^*\zeta_b+\zeta_b^*\zeta_a)/2$, as the normalized eigenvector corresponding to its largest eigenvalue~\cite{Mueller:2004}. Such magnetic-phase mixing can appear, for example, in spinor vortices where atoms in one magnetic phase accumulate at the singular core of a vortex in another phase~\cite{Sadler:2006,Lovegrove:2014,Lovegrove:2016,Weiss:2019}. Analogous excitations have also been studied in superfluid $^{3}\mathrm{He}$~\cite{Finne:2006}.

We follow the experimental procedure outlined in Ref.~\cite{Hall:2016} to create quantum knots in a $^{87}\mathrm{Rb}$ condensate. In brief, we confine $N=2.5\times10^5$ atoms in the $\ket{0}$ spin state in a 1064-$\mathrm{nm}$ crossed-beam optical dipole trap with radial and axial trap frequencies $(\omega_r,\omega_z)=2\pi\times(130,170)~\mathrm{Hz}$. The knot is created by rapidly placing the zero of a three-dimensional quadrupole magnetic field into the center of the condensate and holding it there for $500~\mathrm{\mu s}$, during which time the nematic directors precess into the knot configuration (see Supplementary material for experimental details). %(see Supplemental material at [URL will be inserted by publisher] for experimental details.)

After the knot is created, we eliminate the magnetic quadrupole contribution while rapidly turning on a uniform bias field to $B_0 \simeq 1~\mathrm{G}$ for a subsequent evolution time $T$. After this evolution, we rapidly increase $B_0$ to a large value, after which we release the condensate from the optical trap, separate the spinor components by briefly applying an inhomogenous magnetic field, and image the condensate.

\begin{figure*}[ht]
\includegraphics[width=\textwidth]{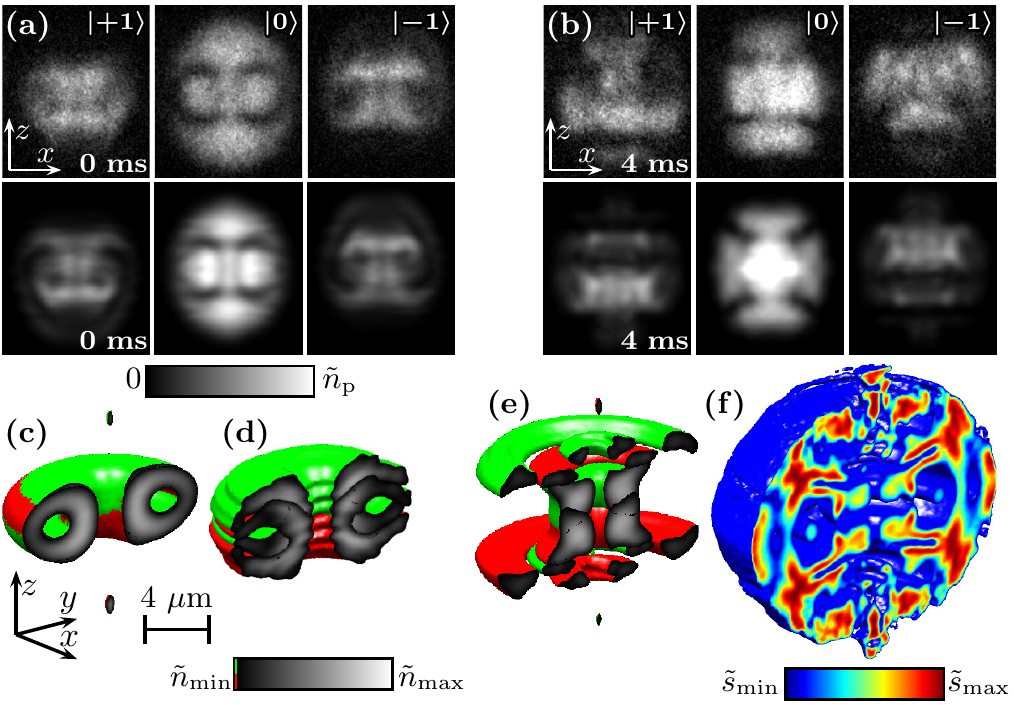}
\caption{(a),(b) Post-expansion column particle densities of the three spinor components integrated along $y$ from experiments (top row) and simulations (bottom row) showing the decay of the knot for evolution times (a) $T=0~\mathrm{ms}$ and (b) $4~\mathrm{ms}$. 
(c)--(e) In-trap density isosurfaces showing the decay dynamics of the initially overlapping hollow vortex rings in $\zeta_{+1}$ (red) and $\zeta_{-1}$ (green) components associated with the quantum knot. The evolution times are (c) $T=0~\mathrm{ms}$, (d) $1~\mathrm{ms}$, and (e) $4~\mathrm{ms}$. 
(f) The spin density isosurface at $T=4~\mathrm{ms}$. 
In each of the panels in (a) and (b), the peak particle density is $\tilde{n}_\mathrm{p}=8.5\times10^8~\mathrm{cm}^{-2}$ and the field of view is $230\times270~\mathrm{\mu m}^2$. 
The particle density isosurfaces in (c)--(e) correspond to $\tilde{n}_\mathrm{min}=3.5\times10^{13}~\mathrm{cm^{-3}}$ with the maximum densities (c) $\tilde{n}_\mathrm{max}=2.6\times10^{14}~\mathrm{cm}^{-3}$, (d) $3.2\times10^{14}~\mathrm{cm^{-3}}$, and (e) $4.4\times10^{14}~\mathrm{cm^{-3}}$. 
The normalized spin density isosurface corresponds to $\tilde{s}_\mathrm{min}=0.5$ with the gradient-maximum $\tilde{s}_\mathrm{max}=1$.}\label{fig:short_evolution}
\end{figure*}

\emph{Results.}---Figure \ref{fig:schematics}(a) shows the numerically determined spinor density isosurfaces for the quantum knot. The column particle densities of different spinor components during the early evolution times of the knot are shown in Fig.~\ref{fig:short_evolution}(a),(b) and the corresponding simulated in-trap particle density isosurfaces of the $\zeta_{\pm1}$ components in Fig.~\ref{fig:short_evolution}(c)--(e). At $T=0~\mathrm{ms}$, the knot structure is visible [see Fig.~\ref{fig:schematics}(a) for reference] with slight displacements in the $\zeta_{\pm1}$ components introduced by the weak magnetic field gradient present during the introduction of the 1-$\mathrm{G}$ bias field and also by the detection process~\cite{Hall:2016}. At $T=1~\mathrm{ms}$, we observe that the initially overlapping $\zeta_{+1}$ and $\zeta_{-1}$ components begin to separate along the negative and positive $z$ axes, respectively. At $T=4~\mathrm{ms}$, the $\zeta_{\pm1}$ components are further displaced from the initial configuration and move to the boundary regions of the condensate. This indicates that $\hat{\bf d}\neq \hat{\bf z}$ at the boundary, and topologically the structure is no longer a pure quantum knot. 

\begin{figure*}[ht]
\includegraphics[width=\textwidth]{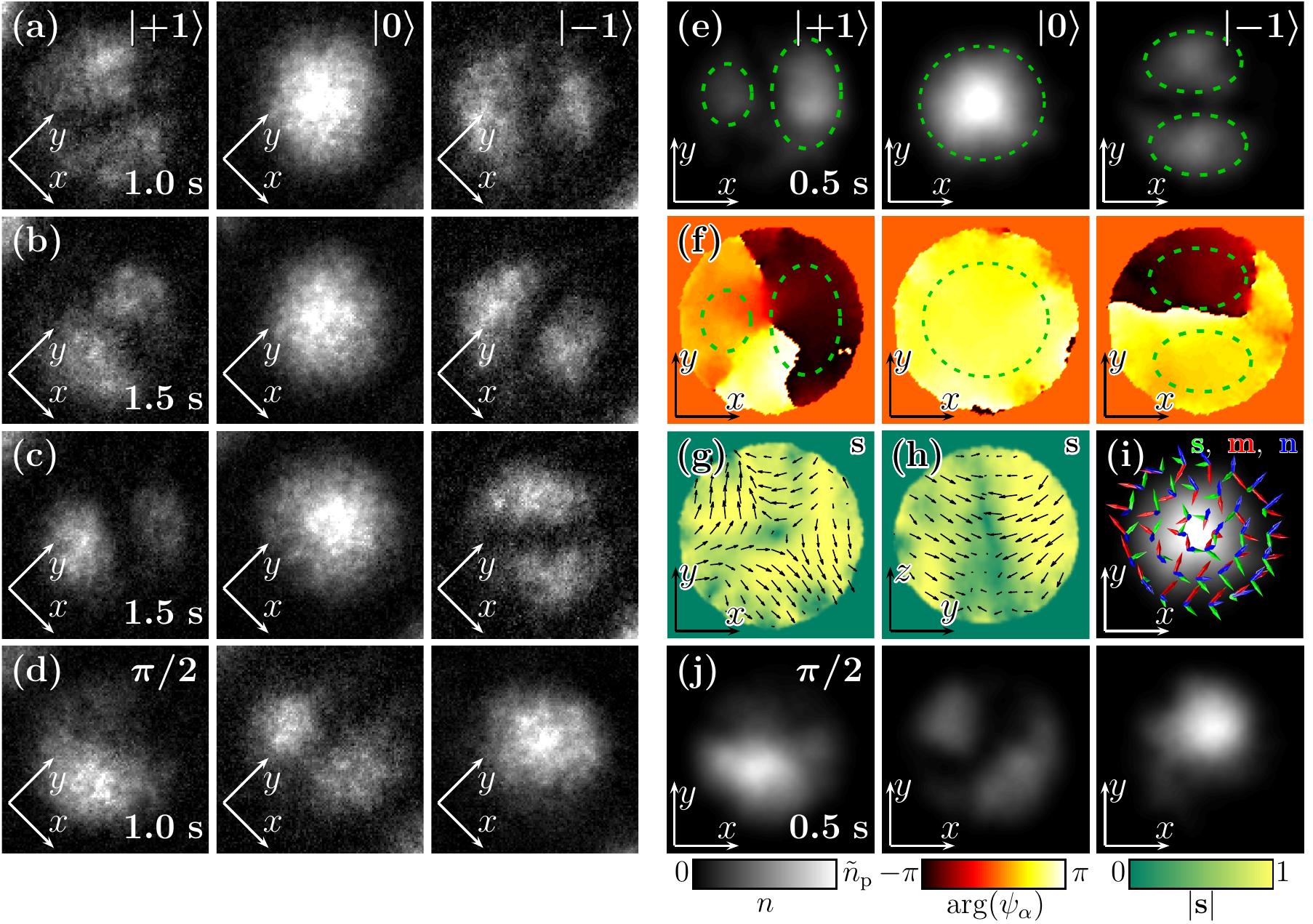}
\caption{(a)--(c) Experimental column particle densities along $z$ of different spinor components of the polar-core spin vortex emerging from the evolution of the quantum knot with the evolution time (a) $T=1.0~\mathrm{s}$ and (b),(c) $1.5~\mathrm{s}$. (e),(f) Simulated particle densities integrated along $z$ and phases in the $z=0$ plane of different spinor components with $T=0.5~\mathrm{s}$ evolution time. The dashed circular shapes are guides for the eye towards the regions with high particle densities. (g),(h) Expectation value of spin, ${\bf s}=\zeta^\dagger{\bf F}\zeta$, in the (g) $z=0$ and (h) $x=0$ planes, with the arrows depicting its planar projection and the color denoting the magnitude $|{\bf s}|$. (i) Triad representation of the order parameter in the $z=0$ plane, where ${\bf s}$, ${\bf m}$, and ${\bf n}$ vectors are represented with green, red, and blue arrows, respectively. The column particle density of $\zeta_0$ is shown for reference. (d) Experimental and (j) simulated column particle densities integrated along the $z$ axis in a $\pi/2$-rotated spinor basis with evolution times $1.0~\mathrm{s}$ in experiments and $0.5~\mathrm{s}$ in simulations. The particle densities in (j) are obtained by expressing the spinor in (e) in $x$-quantized basis. The peak particle density is $\tilde{n}_\mathrm{p}=1.0\times10^9~\mathrm{cm}^{-2}$ and the fields of view of each panel in (a)--(g),(i),(j) are $225\times225~\mathrm{\mu m}^2$ and in (h) $225\times270~\mathrm{\mu m}^2$.}\label{fig:longtime}
\end{figure*}

The spin density isosurface at $T=4~\mathrm{ms}$ [Fig.~\ref{fig:short_evolution}(f)] demonstrates the appearance of ferromagnetic domains early in the evolution. This is partly due to the winding of the $\hat{\bf d}$ vector associated with the knot structure, which gives rise to an inherent instability of the polar phase~\cite{Kawaguchi:2008}. We have numerically verified that there is no discernible difference in the ferromagnetic domain formation between knots created with instantaneous and experimental creation ramps. However, shortly after $T=0~\mathrm{ms}$, we begin to observe some differences between the experimental and simulated particle density distributions, most notably in the $\zeta_{\pm1}$ components [Fig.~\ref{fig:short_evolution}(b)]. We suspect that uncontrolled magnetic fields arising from, e.g., eddy currents induced in nearby metallic structures during the rapid field changes, may play a role in the differences we observe for the early evolution. 

For time scales between $4~\mathrm{ms}$ and several hundred milliseconds, the ferromagnetic and polar regions become intricately scrambled. Surprisingly, at $T\gtrsim 500~\mathrm{ms}$, an emergent polar-core spin-vortex is observed and remains visible for evolution times up to several seconds. Figure~\ref{fig:schematics}(b) shows schematically the spinor structure based upon the agreeing experimental and simulated particle densities shown in Figs.~\ref{fig:longtime}(a)--(c) and (e), respectively. The observed spinor structure is approximately cylindrical, with the polar phase present on the symmetry axis. Away from the axis, the condensate enters into a mixture of polar and ferromagnetic phases that tends towards the the ferromagnetic phase near the boundary. Spontaneously emerging polar-core spin vortices have been predicted in the long-time evolutions of an isolated monopole~\cite{Tiurev:2016_2} and in the absence of any topological excitation in the polar phase~\cite{Saito:2006}. However, the external magnetic field, and thus the resulting quadratic Zeeman shift which tends to stabilize the polar phase, are absent in both of these studies.

The numerically obtained spin texture of the polar-core spin vortex is shown in Fig.~\ref{fig:longtime}(g),(h). The polar core is visible as a depleted spin density along the $z$ axis whereas the spin vector displays a quadrupolar $2\pi$ rotation about $z$. In Fig.~\ref{fig:longtime}(i), we show that the ${\bf s}$ and ${\bf m}$ vectors undergo quadrupolar windings of $2\pi$ about the nonwinding ${\bf n}$ vector along a path enclosing the core in the $z=0$ plane. Similar winding occurs also in the other planes with constant $z$. Thus, the observed spin vortex belongs to the family of singly quantized singular $\mathrm{SO(3)}$ vortices~\cite{UedaBook,Lovegrove:2012}. The spin vector lies mostly in the transverse plane, but tilts slightly towards positive and negative $z$ near the condensate boundary in the regions where $\zeta_{+1}$ and $\zeta_{-1}$, respectively, predominate over the other components. We have numerically verified the absence of mass flow about the vortex core. 

The locations of the regions in which $\zeta_{+1}$ or $\zeta_{-1}$ predominates are observed to change between experimental runs, breaking the cylindrical symmetry of the initial quantum knot. This suggests that fixed residual magnetic field gradients do not drive the evolution. Three examples of the observed polar-core spin vortices with spatially different spinor density distributions are shown in Fig.~\ref{fig:longtime}(a)--(c). 

We provide additional evidence of the presence of a polar-core spin vortex by showing the spinor components in a $\pi/2$-rotated basis in Fig.~\ref{fig:longtime}(d),(j). We implement this rotation experimentally by applying a resonant $7$-$\mathrm{\mu s}$ RF $\pi/2$ pulse within the $F=1$ spin manifold, while in the simulations, we represent the spinor in the $x$-quantized basis by directly applying a $\pi/2$-rotation about the $-y$ axis. In the new basis, the region occupied by the $\zeta_0$ component indicates where, prior to the rotation, the spin pointed approximately perpendicular to the new quantization axis, while the $\zeta_{+1}$ ($\zeta_{-1}$) component indicates the region where the spin was roughly parallel (antiparallel) to the new quantization axis. We find good agreement between the simulations and the experiment, and note that, in the rotated basis, the $\zeta_0$ component does not appear at the center of the condensate for any rotation axis in the $xy$ plane, as one would expect for a $\pi/2$ rotation of $\hat{\bf d}\parallel \hat{\bf z}$.

In our experimental and numerical studies on simple mixtures of spinor components we have not observed a spin vortex to emerge from the evolution. This suggests that some kind of topological defect or otherwise nontrivial spinor structure is apparently required to initiate the dynamics that lead to the polar-core spin vortex in a 1-G magnetic field inducing a significant quadratic Zeeman shift. The emergence of the spin vortex from the quantum knot bears a resemblance to the topological-defect crossing studied in Refs.~\cite{Borgh:2012,Borgh:2013} where different topological defects continously connect through a spatial interface between the magnetic phases. In the present study, however, the crossing from the quantum knot to the spin vortex occurs temporally rather than through a spatial interface.

\emph{Discussion.}---We have observed the decay of a quantum knot, driven by the decay of the polar phase to the ferromagnetic phase. On the time scale of $500~\mathrm{ms}$, after an uncontrollable scrambling of the spinor components, a surprisingly long-lived and apparently stable singular $\mathrm{SO(3)}$ spin vortex emerges. Interestingly, the observed topological transition changes the topological classification of the defect from the third to the first homotopy group, which is allowed by the finite system size. The apparent stability of the spin vortex could be related to a dissipative process by which the minimization of the condensate energy brings the polar atoms together at the core, with the topologically protected spin vortex remaining outside. %The apparent stability of the spin vortex could be related to a possible pinning mechanism introduced by the polar atoms in the vortex core, acting as a kind of plug potential stabilizing the spin vortex. 

Our work demonstrates the rich physics of the dynamics of topological defects in spinor gases. Identifying the exact mechanisms behind the apparent stability of the polar-core spin vortex and the cause for its emergence from the decay of both the isolated monopoles and quantum knots inspire further research. 

We thank Y. Xiao for experimental assistance and M. O. Borgh, J. Ruostekoski, and K. Tiurev for discussions.
We acknowledge funding 
by the Academy of Finland through its Centre of Excellence Program (Grant No. 312300), 
by the European Research Council under Consolidator Grant No. 681311 (QUESS), 
by the Emil Aaltonen Foundation, and 
the NSF (Grant Nos. PHY-1519174 and PHY-1806318). 
CSC--IT Center for Science Ltd. (Project No. ay2090) and 
Aalto Science-IT project are acknowledged for computational resources. 

\bibliography{knot_dynamics}
\bibliographystyle{apsrev4-1}

\end{document}